\documentclass[a4paper,12pt]{article}
\usepackage{overcite}
\newenvironment{figlist}{\begin{list}{\bf Figure \arabic{fig}:}
{\usecounter{fig}\setlength{\labelwidth}{0.5cm}
\setlength{\labelsep}{0.2cm}\setlength{\parsep}
{1ex plus0.2ex minus0.1ex}\setlength{\itemsep}{1ex plus0.2ex}}}
{\end{list}}
\newcounter{fig}

\newcounter{tab}

\begin{document}
\title{Study of the Depolarized Light Scattering Spectra of Supercooled 
Liquids by a Simple Mode-Coupling Model}
\author{
V.Krakoviack\thanks{e-mail : krako@cpma.u-psud.fr} ,
C.Alba-Simionesco\thanks{e-mail : chalba@cpma.u-psud.fr} , M.Krauzman\\
{\small\em  CPMA\thanks{fax-number : (33/0) 1.69.15.42.00} , b\^{a}timent 490,
universit\'{e} Paris-Sud, F-91405 Orsay, France}}
\date{}
\maketitle

\begin{abstract}
By using simple mode coupling equations, we investigate the depolarized light
scattering spectra of two so-called "fragile" glassforming liquids, salol 
(phenylsalicylate) and CKN (Ca$_{0.4}$K$_{0.6}$(NO$_3$)$_{1.4}$), measured by
Cummins and coworkers.
Nonlinear integrodifferential equations
for the time evolution of the density-fluctuations autocorrelation functions
are the basic input of the mode coupling theory. Restricting ourselves to a 
small set of such equations,
we fit the numerical solution to the experimental spectra.
It leads to a good agreement between model and experiment, which allows us to
determine how a real system explores the parameter space of the model, but it
also leads to unrealistic effective vertices in a temperature range where the
theory makes critical asymptotic predictions.
We finally discuss the relevance and the range of validity of these universal
asymptotic predictions when applied to experimental data on supercooled liquids.
\end{abstract}
\vspace{6.5cm}
\pagebreak
\section{Introduction}
\indent

Understanding the properties of the supercooled liquids is still a
challenging problem. This explains the considerable
amount of work on the subject from the experimental as well as from the
theoretical point of view. Of great interest is the fact that, whereas the
static properties of a supercooled liquid vary weakly in the temperature
range between the melting temperature $T_m$ and the so-called calorimetric
glass transition temperature $T_g$, the dynamical properties such as
transport coefficients exhibit a strong temperature dependence. In
particular, for 'fragile' liquids (according to Angell's classification
\cite{ang}), the viscosity shows a strongly non arrhenian behavior : it
increases slowly with decreasing temperature in the weakly supercooled
domain (near $T_m$), but changes by many orders of magnitude when approaching
$T_g$. Below this point, the systems falls out of metastable equilibrium
and becomes a non ergodic amorphous solid, the glass. Such an observation
leads to assume the existence of two dynamical regimes crossing over at a
certain temperature.

From this experimental background, the mode coupling theory (MCT) \cite{mct}
proposes a mechanism for the dynamics of supercooled liquids based on
density fluctuations : a closed set of integrodifferential equations 
is established within the formalism of the
generalized Langevin equation by approximating the memory function as 
polynomials of density correlators;
a bifurcation scenario is then elaborated and describes
the liquid-glass transition as one from an ergodic liquid state with zero
Debye-Weller factor (DWF) to a non-ergodic ideal glassy state with non-zero
DWF. A dynamical critical transition at which the
singularity is expected to occur is then introduced at a temperature $T_{c}$
higher than $T_{g}$.
Due to its critical nature the MCT makes a number of universal predictions 
in the vicinity of $T_c$.
In particular, the $q$-dependence of the density correlators should vanish
in the intermediate-time or -frequency domains where two relaxations, $\alpha$
and $\beta$, overlap,
the dynamics being described by a single function only depending on 
a unique parameter $\lambda$. A square-root singularity of the
DWF and a power-law divergence of the viscosity are predicted too.

To date most comparisons of the MCT to experimental data rely on these
predicted universal features, although the range of validity of the 
approximations leading to these predictions is not well determined
(only recently, the necessary corrections to these asymptotic predictions have
been investigated for hard spheres \cite{correc}).
A different approach is to choose a reduced set of mode-coupling equations
(then called "schematic equations") and to fit its full solution to
experimental data with no direct reference to the expected critical
properties\cite{albkr}.
Such a method, although very simplified, has the advantage of taking
explicitly
into account the microscopic dynamics, which is impossible by other means,
and of allowing an assessment of the range of validity of the universal
predictions.
A restricted study has already been presented for the Raman spectra of
several molecular glassformers ($m$-toluidine and other disubstitued benzenes,
glycerol) in the frequency range from 150 GHz to 6000 GHz \cite{albkr, albkr2}.

In this paper, we propose to apply this method to a wider frequency range 
\cite{pise} in order to fit the depolarized light scattering (DLS) spectra of
salol (an aromatic molecular liquid) and CKN (a binary molten salt)
between 0.3 GHz and 5000 GHz, as measured by Cummins and coworkers.
Many different experimental techniques have been applied
to them, allowing tests of many aspects of the theory. A summary of these
experimental results is proposed for comparison with our results
in table\ \ref{salpast} for salol and table\ \ref{cknpast} for CKN.
In the next section, we briefly describe the experimental support and
some spectral features that our model has to take into account.
In section\ \ref{sec-mct}, we review theoretical results needed for
data analysis and discussion.
Our model and data processing are presented in section \ref{sec-model},
whereas fitting results are exposed in section \ref{sec-results}. A 
discussion of the MCT predictions is given in section
\ref{sec-sceq} and conclusions appear in section \ref{sec-conc}.

\section{Experimental support and spectral trends}
\label{sec-exp}
\indent

We have analyzed the VH light scattering spectra of supercooled salol and
CKN obtained by Cummins and coworkers by combining both Raman and
Brillouin scattering in the frequency range between 0.3 GHz and 5000 GHz.
The temperatures of experimentation cover the whole domain of the supercooled
liquid state, from just above the melting point to the calorimetric glass
transition.
These spectra have been already extensively studied in the context
of the universal predictions of the MCT (\cite{cum1,cum2} for salol,
\cite{cum2,cum3} for CKN) and the reader is referred to these articles
for experimental details.

Rather than the experimental scattered intensity, the susceptibility spectra
are more suited for a comparison with MCT predictions. These are obtained
by normalizing the experimental intensities by the Bose factor :
\begin{equation}
\label{chisec}
\chi''(\omega)= \frac{I^{exp}(\omega)}{n(\omega)+1} \mbox{\ \ with\ \ }
n(\omega)=\frac{1}{\exp(\frac{\hbar\,\omega}{k_B\,T})-1}.
\end{equation}
In the following, the term spectra will always refer to this 
frequency-depen\-dent susceptibility.

The spectra of salol present some general features shared by many
aromatic fragile supercooled liquids of the same kind like 
$o$-terphenyl \cite{pat1, pat2},
$m$-toluidine \cite{albkr}, $m$-cresol and others \cite{albkr2}. These are :
\begin{itemize}
\item a double high-frequency peak with principal frequency at
about 3600 GHz and a shoulder at 600 GHz, often referred to as `Boson' peak
and whose physical interpretation is not yet 
definitive \cite{albkr2, pat2, jac, sok, duv}.
This part of the spectra is almost temperature-independent and is
assigned to microscopic excitations of the system.
\item a low frequency peak which maximum frequency decreases
rapidly with temperature according to an non arrhenian law. It corresponds
to the $\alpha$-relaxation processes.
\item the intensity of the $\alpha$-peak is a lot greater than that of
the microscopic band.
\item the shape of the $\alpha$-peak is characteristic of a stretched, rather
than exponential, relaxation.
\item a minimum occurs between these two structures with an enhancement of
intensity at higher frequencies due to the so-called fast $\beta$-relaxation :
this region, where $\alpha$- and $\beta$-relaxations cross, is the domain of
interest to test the universal predictions made by the MCT. 
\end{itemize}

The spectra of CKN, a fragile ionic glassformer, exhibit the same
features except for two points :
\begin{itemize}
\item  the height of the $\alpha$-peak is much lower, {\em i.e.} it has
roughly the same intensity as the microscopic band.
\item no pronounced extra-structure is distinguishable in the microscopic
region.
\end{itemize}

These differences may originate from the fact that 
both the translational and rotational molecular motions are detected, as 
suggested first for the $o$-terphenyl spectra \cite{pat1}. Indeed, the nature
of the process that dominates VH spectra, interaction-induced or 
reorientation-induced scattering, has been investigated for the different
parts of the spectra of salol \cite{cum5}. It
was found that below the microscopic band the second
mechanism is dominant, while a constant depolarization ratio of 0.75 was 
measured for all these liquids above 600 GHz, supporting a dipole-induced 
dipole (DID) mechanism. Therefore, the observed differences could be
simply related to a different weighting of these two contributions.
This problem will be further discussed in section \ref{sec-results}.

\setcounter{equation}{0}
\section{Mode-coupling theory of the glass transition}
\label{sec-mct}
\indent

As a starting point, the Mode Coupling Theory of the liquid-glass transition
assumes that the dynamics of a supercooled liquid is entirely 
governed by the density fluctuations and proposes a set of generalized kinetic
equations ({\em\`a la} Mori-Zwanzig) for the normalized autocorrelation 
functions $\phi_{q}(t)$ of the density fluctuations modes at wave vector $q$,
$\rho_{q}(t)$ \cite{mct}.
These equations take the form :
\begin{equation}
\ddot{\phi}_{q}(t)+\Omega_{q}^{2}\,\phi_{q}(t)+\int^{t}_{0}M_{q}(t-t')\,
\dot{\phi}_{q}(t')\,dt'=0,  
\end{equation}
where the memory kernel is split
into an instantaneous damping term $ \gamma_{q}\,\delta(t) $ and a 
mode-coupling term $m_{q}(t)$ :
\begin{equation}
M_{q}(t) = \Omega_{q}^{2}\,[\gamma_{q}\,\delta(t)+m_{q}(t)].
\end{equation}
The core of the mode coupling approximation in the "idealized case" only 
considered here is to express $m_{q}(t)$ as a 
polynomial function $F_{q}$ of the $\phi_{q}(t)$'s, so that
\begin{equation}
\label{memfun}
m_{q}(t)= F_{q}({\mathbf V},\phi_{k}(t))=\sum_{n \geq 1}\  
\sum_{q_{1},\ldots,q_{n}} V_{q_{1},\ldots,q_{n}}\,
\phi_{q_{1}}(t)\,\ldots\,\phi_{q_{n}}(t),
\end{equation}
where the components of the vector ${\mathbf V}$,
the vertices $V_{q_{1},\ldots,q_{n}}$, are positive continuous functions
of the static structure factor $S(q)$. It should be noticed that the linear 
terms in $m_{q}(t)$ are introduced in an {\em ad hoc} fashion to guarantee
that models with only a very reduced set of equations reproduce
the stretched behavior of the $\alpha$-relaxation. Some 
justifications of the presence of these terms have been proposed in the 
context of the non-linear fluctuating hydrodynamics \cite{maz}.
The mode-coupling equations are finally written as :
\begin{equation}
\label{mcteq}
\ddot{\phi}_{q}(t)+\gamma_{q}\,\Omega_{q}^{2}\,\dot{\phi}_{q}(t)
+\Omega_{q}^{2}\,\phi_{q}(t)+\Omega_{q}^{2}\,\int^{t}_{0}\!m_{q}(t-t')
\,\dot{\phi}_{q}(t')\,dt'=0,   
\end{equation}
with $m_{q}(t)$ given by equation \ref{memfun}.

This set of equations can lead to an ergodic-to-non ergodic transition.
Indeed, the Debye-Weller factors (DWF) defined by 
\begin{equation} 
f_{q}=\lim_{t\rightarrow + \infty} \phi_{q}(t)
\end{equation}
are solutions of the set of equations :
\begin{equation}
\label{dwf}
\frac{f_{q}}{1-f_{q}}= F_{q}({\mathbf V},f_{k}).
\end{equation}
The only real solution in the vicinity of ${\mathbf V}\!\!=\!0$ is the 
trivial one (for all $q$'s,$\,\, f_{q}=0$) corresponding to an ergodic liquid 
state. 
A bifurcation to a non zero solution corresponding to a non ergodic "ideal 
glassy state" occurs when a critical hypersurface, the "ideal glass 
transition hypersurface", is crossed. It is defined by a vector 
${\mathbf V}^{c}$ such that the 
Jacobian matrix of equations \ref{dwf} is singular, so that :
\begin{equation}
\frac{f^{c}_{q}}{1-f^{c}_{q}}= F_{q}({\mathbf V}^{c},f^{c}_{k})
\end{equation}
\begin{equation}
\label{matrix}
\det \left[ \frac{1}{(1-f^{c}_{q})^{2}}\,\delta_{qp}-\frac{\partial F_{q}}
{\partial f_{p}}({\mathbf V}^{c},f^{c}_{k}) \right] =0.
\end{equation}
Two type of transitions exists : the A-type with the DWF varying continuously
through the transition hypersurface and the B-type with discontinuous change
of the DWF.
The second one is the only potentially relevant in the context of the 
liquid-glass transition. At the thermodynamic level, intensive critical 
parameters (critical temperature, density and so on), at which the 
transition is supposed to occur, are defined. 

In the vicinity of the transition hypersurface, it is possible to make
various predictions on the critical behavior of the solutions of the mode
coupling equations, the most important being the "reduction theorem".
Expanding to lowest order in ${\mathbf V}-{\mathbf V}^{c}$, one obtains
\begin{equation}
\label{expansion}
\phi_{q}=f^{c}_{q}+(1-f^{c}_{q})^{2}\,G_{q}(t),
\end{equation}
where the leading contribution to $G_{q}$, in the vicinity
of the plateau of $\phi_{q}$, is proportional to a $q$-independent
function $G$, which is solution of the equation (called the scaling equation)
\begin{equation}
\label{eq:scaling}
\sigma + \lambda\,G(t)^2 - \frac{d}{dt}\int^{t}_{0}G(t-t')\,G(t')\,dt'=0,
\end{equation}
where $\sigma$, the separation parameter, is an $O({\mathbf V}
-{\mathbf V}^{c})$ given by
\begin{equation}
\sigma=\sum_{q} \hat{e}_{q}\,\left(F_{q}({\mathbf V},f^{c}_{k})
-\frac{f^{c}_{q}}{1-f^{c}_{q}}\right)   
\end{equation}
and is positive in the ideal glassy state, negative in the liquid, and zero on
the transition hypersurface. The exponent parameter $\lambda$ is defined by
\begin{equation}
\lambda = \frac{1}{2}\,\sum_{q,k,p} \hat{e}_{q}\,\left( \frac{\partial^{2}
F_{q}}{\partial f_{k}\, \partial f_{p}}({\mathbf V}^{c}, f^{c}_{l})\right)
\,(1-f^{c}_{k})^{2}\,(1-f^{c}_{p})^{2}\,e_{k}\,e_{p},
\end{equation}
with $[e_{q}]$ and $[\hat{e}_{q}]$ eigenvectors of the matrix 
\ref{matrix} \cite{mct}.

Eqs.\ \ref{expansion} and \ref{eq:scaling} and some of their consequences have
been extensively used to test the validity of MCT by comparison with 
experimental data. 
In particular, an approximate expression around the susceptibility minimum
has been proposed :
\begin{equation}
\label{approx}
\chi''(\omega)=\frac{\chi''_{min}}{a+b}\,\left(b\,\left(\frac{\omega}
{\omega_{min}}\right)^{a} + a\, \left(\frac{\omega_{min}}{\omega}\right)^{b}
\right),
\end{equation}
with $a$ and $b$ such that
\begin{equation}
\lambda=\frac{\Gamma(1-a)^{2}}{\Gamma(1-2\,a)}=\frac{\Gamma(1+b)^{2}}{\Gamma
(1+2\,b)}.
\end{equation}
The relevance of such studies will be discussed in section \ref{sec-sceq}.

\setcounter{equation}{0}
\section{Model and data processing}
\label{sec-model}
\indent

In our calculations, we intend to take into account the properties
of the spectra in the THz domain such as the microscopic peak and the 
inelastic boson peak.
This is not possible with studies by means of the scaling equation, eq.
\ref{eq:scaling} or eq. \ref{approx}, because they are
restricted to the region around the minimum of the susceptibility spectrum,
{\em i.e.}  to the GHz domain. 
Thus, we choose to study in full detail a schematic model, {\em i.e.} a 
restricted set of mode-coupling equations like eq. \ref{mcteq} with a
simple form of the memory functions, eq. \ref{memfun}.

The basis of our study is the well-studied $F_{12}$-model \cite{mct} for a
single correlator $\phi_{0}$ defined by
\begin{equation}
\label{phi0}
\ddot{\phi}_{0}(t)+\Gamma_{0}\,\Omega_{0}\,\dot{\phi}_{0}(t)
+\Omega_{0}^{2}\,\phi_{0}(t)+\Omega_{0}^{2}\,\int^{t}_{0}\!m_{0}(t-t')
\,\dot{\phi}_{0}(t')\,dt'=0,   
\end{equation}
\begin{equation}
\label{mem0}
m_{0}(t)= v_{1}\,\phi_{0}(t)+v_{2}\,\phi^{2}_{0}(t).
\end{equation}
We add a second correlator $\phi_{1}$ in order to reproduce the double-hump
shape of the peak in the THz domain and the extra-intensity of the 
$\alpha$-peak. It is solution of
\begin{equation}
\label{phi1}
\ddot{\phi}_{1}(t)+\Gamma_{1}\,\Omega_{1}\,\dot{\phi}_{1}(t)
+\Omega_{1}^{2}\,\phi_{1}(t)+\Omega_{1}^{2}\,\int^{t}_{0}\!m_{1}(t-t')
\,\dot{\phi}_{1}(t')\,dt'=0.   
\end{equation}
It remains to express the memory kernel $m_{1}$. For this purpose, we take
the first correlator $\phi_{0}$ accounting in an effective way for 
all modes that are responsible for the underlying critical transition 
(typically, these are thought to be the density fluctuations at wave vectors
corresponding to the maximum of the static structure factor \cite{geszti}). 
The second correlator $\phi_{1}$ describes additional degrees 
of freedom, that are important in depolarized light scattering processes
but whose slow time-dependence is dominated by that of $\phi_{0}$ (so
that the liquid-glass transition occurs systematically for the two correlators
at the same time). A simple choice is then
\begin{equation}
\label{mem1}
m_{1}(t)=r\,m_{0}(t).
\end{equation}
In addition we impose that the characteristic pulsations ($\Omega_{0}$,
$\Omega_{1}$), the damping coefficients ($\Gamma_{0}$, $\Gamma_{1}$)
and the coupling parameter $r$ are temperature-independent. As a consequence,
the whole temperature dependence is contained in the vertices $v_{1}$ and 
$v_{2}$.

The above model has a B-type transition, and the critical hypersurface reduces
to a line in the 2-dimensional parameter space ($v_{1}$, $v_{2}$) :
\begin{equation}
v^{c}_{1}=2\,\sqrt{v^{c}_{2}}-v^{c}_{2} \mbox{\ \ \ with \ \ } 
1 \leq v^{c}_{2} \leq 4 
\end{equation}
The exponent parameter is :
\begin{equation}
\lambda=\frac{1}{\sqrt{v^{c}_{2}}}.
\end{equation}

Other models with two correlators have already been proposed in a similar 
context : the Sj\"{o}gren model for diffusion of impurities
in a glass-forming liquid \cite{sjo} and the 
Bosse-Krieger model for coupled mass-density and charge-density correlators 
in a supercooled molten salt \cite{bosse}.  

It should be noted that, because of the extreme simplicity of the model, 
the real nature of the correlators is somewhat obscured.
$\phi_{0}$ and $\phi_{1}$ are not necessarily restricted to be density 
correlators. In particular, the model could account for molecular 
orientional fluctuations or any other degrees of freedom that are not 
included in standard MCT. Accordingly, the parameters that we extract from
the fits do not have a precise microscopic content and should be considered 
as a minimal set of phenomenological quantities that allows to describe the 
global evolution of the systems in the context of simple mode-coupling models. 

To compare the predictions of the model to the experimental data, it is also 
necessary to make some assumptions about the mechanism that gives rise to 
the light scattering spectra.
When the dipole-induced dipole (DID) mechanism dominates, as often 
assumed \cite{cum2,cum6}, the Stephen factorization scheme \cite{steph}
(which is compatible with the factorization approximations underlying the MCT)
leads to an expression for the DLS spectrum $\chi''(\omega)$ which involves the
Fourier transform of bilinear products of the correlators.
A simple description of the spectrum for the case of our two-correlator model
is similarly given by :
\begin{equation}
\label{spec}
\chi''(\omega)=\omega\,A\,\,\mbox{Im}\left\{ \mbox{FT}
\left[\phi_{0}(t)^{2}+\gamma\,\phi_{1}(t)^{2}\right]\right\}
\end{equation}
with FT denoting Fourier transform and Im the imaginary part.
$A$ is an amplitude factor that we take dependent of temperature.
$\gamma$ is an {\em ad hoc} parameter which specifies the relative weight of
the two contributions to the DLS spectrum due to $\phi_{0}$ and to $\phi_{1}$
and which we take as temperature independent. However, both $A$ and $\gamma$
are substance-dependent.
This expression of the scattered intensity corresponds to the assumption that
$\phi_0$ and $\phi_1$ are both density correlators. Nevertheless, as pointed
in ref. \citen{cum5}, fluctuations of the molecular orientations are a major
contribution to the DLS spectra. In order to test the dependence of our
results in the chosen expression of the scattered intensity,
we have also performed tests with other choices for the expression of
the susceptibility $\chi''(\omega)$, like a linear combination of the
correlators or 'mixed' models, quadratic in $\phi_0$ and linear in $\phi_1$
\cite{soon}.  
The fits were not as good as with eq. \ref{spec} in the microscopic part of 
the spectra, but 
they were equivalent in the region of the minimum of susceptibility and of the 
$\alpha$-peak. The
values of the vertices and of other parameters of the model {\em were only
very weakly affected by the modification}. It seems thus that 
the question of the nature of the light scattering mechanisms is not 
determining for our study. It remains out of our scope.

The calculations consist of three steps :
\begin{itemize}
\item 1) the time evolution of the correlators $\phi_{0}$ and $\phi_{1}$
is computed from equations \ref{phi0} to \ref{mem1} by means of an algorithm
derived from the one described in reference \cite{fuchs} and provided to us
by M.Fuchs and W.G\"{o}tze.
\item 2) The Fourier transform in equation \ref{spec} is
performed via the Filon algorithm \cite{filon} to obtain theoretical
susceptibilities $\chi''(\omega)$.
\item 3) The preceding calculations are included in a loop using 
a modified least-square Powell algorithm \cite{powell} where the 
$T$-dependent parameters $A$,$v_{1}$ and $v_{2}$ and the T-independent 
parameters $\Omega_{0}$, $\Omega_{1}$, $\Gamma_{0}$, $\Gamma_{1}$, $r$ and
$\gamma$ are adjusted to provide the best fit between model and experiment.
\end{itemize}

Analogous calculations have been performed for different choices of
the polynomial approximation to the memory function $m_{0}$ to fit Raman 
scattering spectra of $m$-toluidine \cite{albkr}. 
Good and stable fits were obtained with the $F_{12}$-model associated
with equation \ref{spec} for $\chi''(\omega)$,
so we restrict mostly ourselves to this case. We have used
different functional forms of the memory functions 
to fit the salol spectra, but the results are still preliminary and will be
only briefly mentioned\cite{soon}.

\setcounter{equation}{0}
\section{Numerical results}
\label{sec-results}
\indent 

A major difficulty with the MCT in its formulation only in terms of density
fluctuation modes is that it predicts the existence of a
sharp ergodic-to-non ergodic transition that is not observed 
experimentally. This transition which is assumed to be at a temperature
$T_{c}$ higher than $T_{g}$ is avoided because of so-called activated 
processes that are not taken into account by the theory and that restore 
ergodicity below $T_c$.
For this reason, the low-temperature features predicted below $T_{c}$ are
always obscured in the supercooled liquid state and the rare 
characterizations of these \cite{cum3} do not come without any 
polemics \cite{tarjus}. 
It is thus necessary to impose some temperature limitations to the fitted
data by ignoring the coldest spectra or imposing a low-frequency 
cut-off for the lowest temperatures in order to free our
calculations from this difficulty without loosing any microscopic information. 
It is clear that such constraints will affect the fitting results. This point
is discussed later on.

Typical fits to experimental data are shown in figure \ref{fit}
at different temperatures for both liquids. With the chosen cut-offs
(typically about 100 GHz for temperatures below the best fit to $T_{c}$. See
fig. \ref{fit}.a), the schematic spectra are in very good
 agreement with the data. In particular, at high 
temperature, the whole spectra are reproduced by the model 
including well defined $\alpha$ and microscopic peaks.
Thus this set of equations in the mode-coupling formalism ensures the
continuity of the dynamical processes from the transient vibrational regime
of the microscopic band to the regime of structural relaxation.
A drastic difference of behavior exists between high and low temperatures 
as far as a choice of frequency cut-off is concerned.
Out of the restricted fitting frequency range, a reasonable extrapolation to 
low frequencies is possible for high temperatures, whereas it is impossible 
at low temperatures. This is due to the limitations of the theory 
discussed above. In the first case (high $T$), the spectra
can be described unambiguously by the model, but, in the second case (low 
$T$), the theory does not describe the $\alpha$-relaxation at all and the 
fits force the low-temperature predictions of MCT to occur below the frequency
cut-off, giving unstable and unphysical results.

The calculated temperature-independent parameters of the model
are presented in table \ref{global}. 
The choice of frequency cut-off has little influence on 
these values. Error bars of about 15 \% are determined from different choices.
The values obtained for the frequencies are realistic in regard to
the position of the microscopic bands.

The coupling parameter $r$ measures the strength of the coupling of the
correlator $\phi_{1}$ to the driving correlator $\phi_{0}$. From the fits to
both salol and CKN data, it is found greater than one. This supports the 
assumption that processes can be isolated that drive the overall dynamics of 
the system.

An interesting result is that the relative weight $\gamma$ of the second 
correlator
contribution to the DLS spectra is greater than one and greater for salol
than for CKN. This is illustrated in figure \ref{split} where the two terms 
of equation \ref{spec} are considered separately. It is clear from the figure
that the 
main contribution to the spectra is due to the second correlator $\phi_1$,
except for the highest frequencies. Therefore the first correlator seems to
drive the overall slowing down of the system, 
whereas the second one provides the experimental probe of the dynamics. 
It is an interesting consequence of the fitting procedure chosen with 
eq.\ \ref{spec} to allow such a classification.
Moreover, we may assess the contribution of the second correlator 
depending on the chemical nature of the liquid (aromatic or ionic), at
least with our specific $F_{12}$ model.
Departing from the original MCT scenario that deals only with density 
fluctuations and since our results are independent of the expression of
the scattered intensity as a function of the correlators (see section
\ref{sec-model}), thus of their precise physical content, one could
interpret this result in a way compatible with ref.\ \citen{cum5} :
orientational fluctuations represented by $\phi_{1}$ produce the main part of
the depolarized light scattering spectra and are strongly coupled to
density fluctuations represented by $\phi_{0}$ which drive the dynamics of
the supercooled liquid above $T_c$. It would be necessary to support this
assumption to demonstrate that the dynamics of the orientational and density 
fluctuations obey similar mode-coupling equations, which remains to be done.

We have also let $r$ and $\gamma$ vary with temperature.
However, it does not produce better fits despite the considerable amount of
new adjustable parameters and it was abandoned.

The most important output of our calculations are the effective-temper\-ature
dependent vertices 
$v_{1}$ and $v_{2}$ (figures \ref{v1v2T} and \ref{v1v2}). 
Indeed, the knowledge of these parameters gives access to the way
a realistic system explores the parameters space of the model (see figure 
\ref{v1v2}). It thus allows to discuss the applicability of the asymptotic
predictions of the theory in the vicinity of $T_c$. See below. 

For determining the vertices, the problem of the low-frequency cut-off must be
addressed, again because of the limitations of the theory discussed above. 
At high temperatures, the vertices are determined unambiguously, irrespective
of the cut-off : whether one restricts the fitting range to the microscopic 
peak and its
"fast-$\beta$-relaxation" wing ({\em i.e.} a cut-off at about 100 GHz) or 
includes the region of the susceptibility minimum,
and considers or not the $\alpha$-peak has no influence on the results. 
On the contrary, at low temperatures where the theory cannot apply to the full 
spectrum, 
the fitting procedure gives results that are cut-off-dependent and unstable.
This produces unphysical discontinuities in the behavior of
$v_{1}$ and $v_{2}$ when crossing the dynamical-transition line.
With no frequency cut-off (other than the one due to the experimental 
frequency window), the limitations of the theory appear in a rather
surprising way on the fitted values of the vertices. At the lowest 
temperatures, the
representative curves of salol and CKN seem to follow asymptotically the
transition line as the temperature is lowered (see figure \ref{v1v2}). 
Implications of this observation are discussed in the next section.

Note finally that the choice for the expression of the DLS susceptibility as 
a function of the
correlators $\phi_{0}$ and $\phi_{1}$  has no influence on the values of the
vertices that we obtain. Linear or quadratic expressions produce 
undistinguishable results. This supports our analysis and conclusions, which 
are independent of the detailed description of the light scattering process.

\setcounter{equation}{0}
\section{Critical properties}
\label{sec-sceq}
\indent

Knowing how a real system explores the parameter space of the model can give 
some insight into the applicability of the MCT universal predictions to 
experimental data. This is the purpose of our schematic model to provide such
a knowledge.

As discussed in the preceding section, the "idealized" MCT used here
fails in describing the dynamics of the strongly supercooled liquids,
{\em i.e.} below the critical temperature $T_c$ introduced by the theory 
itself.
It is thus necessary in general to impose temperature and/or frequency 
limitations to the experimental data under study in order to compare with the
theoretical predictions. Fortunately, this has no influence on the
high-temperature results. 
For this reason, we consider in the following only the results obtained 
without explicit frequency cut-off. Doing so, the cut-off dependence of our 
results, a possible objection to mode-coupling tests \cite{tarjus},
is avoided and the comparison with other tests of the theory is made 
easier. The penalty is of course that one must impose a low-temperature
cut-off.

The adjusted effective vertices exhibit two regimes with temperature.
Above and slightly below the melting temperature, they vary linearly
($v_{1}$ is almost constant), whereas, at lower temperature, they tend to 
follow asymptotically the transition line without crossing it.
As the static structure factor evolves only slowly with temperature in the 
supercooled liquid
state, the rapid departure from the quasi-linear behavior appears to signal
a breakdown of the usual MCT hypothesis that the vertices are continuous 
functions of the static correlations and vary moderately with temperature.  
If the critical properties are determined by extrapolation from the
quasi-linear domain only (table \ref{trans}), 
the critical temperatures $T_c$ for salol and CKN are found in good agreement
with the previous determinations (compare table \ref{trans} and tables 
\ref{salpast} and \ref{cknpast}).
It must be noted that the corresponding critical temperatures fall in the
low-temperature but still ergodic regime.
An exponent parameter $\lambda_{HT}$ can be determined from this 
high-temperature behavior. It is in good agreement with previous 
determinations for salol but is significantly smaller for CKN (see tables).

Other extrapolations are possible by including the nonlinear regime. 
Once one includes this regime, the critical temperatures and the exponent 
parameters depend on the low-temperature cut-off, and, if one wants to include
more data, on low-frequency cut-offs as well.
It is found that the smaller the low-temperature cut-off, the smaller the 
extrapolated critical temperature, and the smaller the exponent parameter 
$\lambda$, as can be seen in figure \ref{v1v2}.
In an extreme situation, without any frequency cut-off (and with a wider 
experimental frequency window), because the fitting procedure forces $T_c$ to
be smaller than all the considered temperatures, it appears that no transition
could be defined at all, as the vertices follow asymptotically the transition 
line without crossing it. Unphysical values could then be obtained, like
$v_{1}$ becoming negative. 

Of particular interest is the exponent parameter extrapolated from the 
effective vertices at the critical temperature determined from the linear 
regime.
It is referred to as $\lambda_{vert}$ in table \ref{trans}. 
Whereas one would expect $\lambda_{vert}$ to be compatible with the previous 
determinations of $\lambda$
based on the critical predictions of the theory (tables \ref{salpast} and 
\ref{cknpast}), it is found significantly smaller for both salol and CKN. 
This discrepancy should then affect strongly the quality of a master equation
fit and limit its range of validity.

We have also proceeded with different schematic models, 
a $F_{13}$-based one with
\begin{equation}
m_{0}=v_{1}\,\phi_{0}+v_{3}\,\phi_{0}^{3} \ \ \ \  m_{1}=r\,m_{0} 
\end{equation}
and a Sj\"{o}gren-like one \cite{sjo} with
\begin{equation}
m_{0}=v_{1}\,\phi_{0}+v_{2}\,\phi_{0}^{2} \ \ \ \  m_{1}=r\,\phi_{0}\,
\phi_{1}. 
\end{equation}
Results of the fitting procedure are in total agreement with the above 
picture\cite{soon}.

It is interesting to relate our results to the numerical studies on 
binary mixtures of Lennard-Jones atoms \cite{kob} and soft 
spheres \cite{barrat}.
For both systems, the static structure factors have been used to determine the
vertices of the memory kernels in the mode-coupling equations. Then, the 
critical properties
and the long-time dynamics obtained from the equations 
have been compared to the molecular dynamics simulation results.
It is systematically found that this mode-coupling approach overestimates the
ability of the system to freeze in an "ideal glassy state". The physical 
interpretation of this
effect is clear : restricting the relevant dynamical modes to the density
fluctuations excludes possible alternative decay channels for relaxations and
the freezing occurs more rapidly.
In the case of the Lennard-Jones system, the exponent parameter calculated 
from the mode-coupling model ($\lambda=0.708$; for soft spheres it is 
$\lambda=0.73$) is found smaller 
than that determined from a fit to the scaling equation ($\lambda=0.78 
\pm 0.02$).
For both systems, the diffusion constant seems to vanish (at an extrapolated
temperature) with a power law corresponding to $\lambda\simeq0.60$, whereas 
the $\alpha$-relaxation times seem to diverge with power laws 
corresponding to $\lambda\simeq0.79$ for the Lennard-Jones system and 
$\lambda\simeq0.88$ for the soft spheres. 
This is in contradiction with the requirement that a given system be
characterized by a single exponent parameter $\lambda$.

These discrepancies on the value of $\lambda$ are surprinsingly quantitatively
very similar to those found from our schematic calculations, although 
rigorous and complete sets of mode-coupling equations (and not an arbitrary 
phenomenological model) are considered. Consequently, it seems reasonable to 
us to assume that a schematic study of the molecular dynamics 
simulation results would reveal effective behaviors similar to those found 
for salol and CKN.
 
On the basis of the preceding discussion, we suggest 
a possible description of the approach of the calorimetric glass transition 
in term of effective temperature-dependent mode-coupling parameters.
At high enough temperatures, the dynamics of a liquid
are appropriately described by the ideal mode-coupling theory (in the sense 
that the dynamics are dominated by processes that couple the way the
density correlators do in the MCT with vertices that are smooth functions of
the static correlations in the system).
When the temperature is lowered,
the system seems to evolve toward the ergodic-to-non ergodic transition 
predicted by the theory.
But, at a temperature notably higher than the extrapolated $T_c$, the 
dynamics depart from the predictions of the theory~:
it is better expressed in term of effective vertices that are not directly
related to the statics and behave such that the transition is avoided.
Finally, at low temperature, a qualitative change in the nature of the
relaxation processes reveals a possible underlying dynamical transition.
Quantitatively, this picture exhibits two properties :
\begin{itemize}
\item the ideal mode-coupling approach leads to an overestimated ability of 
the liquids to freeze.
\item incompatible values of the exponent parameters are determined :
compared to the one calculated from the statics, the fitting of the scaling
equation overestimates $\lambda$, whereas the effective behavior leads to
an underestimation.
\end{itemize}
 
If this picture works well for CKN,
this last point is not clear in the case of salol, where our schematic fits
provide equal exponent parameters from the high temperature extrapolation and
the scaling equation.
It is thought to be a peculiarity of this system, when approached with our
elementary model.
In CKN, there is a relative equivalence between the contributions of the two
correlators to the scattered intensity (the proportions are about 30\%-70\%),
whereas in salol the second correlator represents about 95 \% of the scattered
intensity,
as a consequence of the height of the $\alpha$-peak much larger than the
microscopic peak.
Thus our calculations for salol reduce in reality in the GHz domain to a fit 
of this 
single correlator, that is in fact equivalent to a fit of the scaling equation
in this domain (see figure \ref{split}). Only more complex models with more 
correlators could reveal the proposed picture. 
 
The nature of the processes responsible for the transition to an effective
mode-coupling behavior is of course of great interest.
It is not clear that discrete processes as correlated jumps and hopping 
are good candidates, because in simulations they are generally characterized 
only well below the extrapolated critical temperature of MCT (for 
Lennard-Jones mixture \cite{kob}, $T_{c}=0.922$ is predicted from statics, 
whereas the secondary peak of the self part of the van Hove correlation 
function generally associated to these processes appears below 0.5).

If the picture described here is valid, it introduces some questions 
about the relevance of the predictions of MCT in the vicinity of the 
dynamical critical transition, that we found to be widely avoided. 
It is thus necessary to know what is the maximum distance to the critical 
hypersurface at which the universal predictions of the theory are
effective : schematic models can provide some answers to this determining 
problem.

First consider the elementary $F_{12}$-model for one density correlator 
defined by equations \ref{phi0} and \ref{mem0} alone.
Following G\"{o}tze and Sj\"{o}gren \cite{gotze},
we choose
\begin{equation}
v_{1}=v_{1}^{c}\,(1-\varepsilon)\ \ \ \ \ \ \ 
\ \ v_{2}=v_{2}^{c}\,(1-\varepsilon)
\end{equation}
such that $\lambda=1/\sqrt{v_2^{c}}$ equals 0.7 on the transition line.
$\varepsilon$ is thus a kind of 'distance' from the transition line.
After numerical in\-te\-gration and Fourier transform, the susceptibility
spectra are fitted in the minimum region with the scaling equation 
(equation \ref{eq:scaling}) in order to determine an effective exponent 
parameter $\lambda_{e\!f\!\!f}$ as a function of positive $\varepsilon$ 
(figure \ref{lambda}). This procedure reproduces the usual
way of treating experimental results beyond the scope of MCT.
It is clear that for $\varepsilon$ greater than 0.005, $\lambda_{e\!f\!\!f}$
departs notably from its value on the critical line and is never constant. It 
corresponds thus to a maximum distance from the transition line for the 
critical properties predicted by the theory to be effective.

In order to take into account this $\varepsilon$ dependence, a vertices
dependent exponent parameter can be defined following G\"{o}tze and Sj\"{o}gren
\cite{lamb}.
In the case of the $F_{12}$-model and of our schematic model, which have
exactly the same critical properties, this reads :
\begin{equation}
\lambda_{loc}=v_{2}\,(1-f)^{3} \mbox{\ \ with $f$ solution of\ \ }
v_{1}+2\,v_{2}\,f=\frac{1}{(1-f)^{2}}
\end{equation}
valid only if
\begin{equation}
\label{valid}
v_{1}\geq 3\,v_{2}^{2/3}-2\,v_{2}
\end{equation}
We see that, in the $F_{12}$-model, the introduction of local scaling
equation parameters extends the domain of numerical compatibility of this 
equation to $\varepsilon=0.015$, but it is not enough :
effective parameters are determined out of this range and
even in the domain where the condition \ref{valid} is not obeyed.
These results have thus no significance in term of the MCT and  
arise only because of the numerical flexibility of the fitting procedure. 

The same problems should appear with the experimental spectra.
Indeed the trajectories followed by salol and CKN in the parameters space of
our model are not nearer, in term of a distance defined as $\varepsilon$ is,
from the transition line than are the points considered above. 
Moreover some experimental points do not obey the constraint \ref{valid}.

We have performed scaling equation calculations as above for the 
susceptibility spectra of salol between 263 and 333 K. It is clear from 
figure \ref{lambsalol} that the values of $\lambda_{e\!f\!\!f}$ vary notably 
with temperature and can be determined even when \ref{valid} is not valid. 
They are compatible with the previous determinations by other authors 
(table \ref{salpast}) but neither with $\lambda_{vert}$ (table \ref{trans})
nor with $\lambda_{loc}$ coming from our schematic calculations\cite{lamb}.
This is representative of the ineffectivity of the reduction theorem, as is
the existence of two distinct minima for each correlator at all temperatures
(figure \ref{split2}).
Two possible reasons can be proposed for this, depending on the considered 
temperature domain.
At high temperature, where the description of the dynamics 
by MCT is correct and alternative processes are not necessary,
the supercooled liquid is too far from its transition point and thus the 
asymptotic predictions of the theory valid in the vicinity of the transition
do not yet hold, whereas,
at low temperature, processes that are not tractable within the mode-coupling
formalism are probably at the origin of these discrepancies.  

The point is thus that our simple schematic calculations indicate that the
leading order critical predictions of the MCT are experimentally unobservable.
The conclusion would have been the same even if there were no need of 
additional processes to describe the dynamics of the supercooled liquid in 
the transition region; it is found necessary the system to be very 
near its transition point,
where error bars of mosts experimental techniques impose
differences of several degrees to reveal significant evolutions. This necessity
has been pointed out recently within the investigation of the next-to-leading
order corrections to the critical predictions of MCT\cite{correc}, which
are not for the moment available for experimental data like DLS spectra.

\section{Conclusion}
\label{sec-conc}
\indent

We have analyzed depolarized light scattering spectra of two
fragile glassforming liquids, salol and CKN, by means of schematic 
mode-coupling
equations ({\em i.e.} a reduced closed set of non linear integrodifferential 
equations for correlation functions).
Such simple models are interesting because they make possible tests
of the MCT that are not limited to its asymptotic critical predictions.
Moreover, they capture the essential structure of the mode-coupling
equations with no reference to the details of the physical processes at
work. It is thus possible to take into account in an effective manner
processes that have not been included so far in the MCT.
It is the case for instance of the orientational fluctuations.

We have found that the two-correlator model developed here gives a very good 
account of the experimental data. In particular, the so-called boson peak
offers no extra-difficulty within our fitting procedure, whatever functional
form is chosen for the memory kernel \cite{albkr, albkr2}. 

The evolution with temperature of the relaxational properties of the system
can be rationalized in term of 
effective temperature-dependent mode-coupling vertices. It is
compatible under certain assumptions with observations previously made
in molecular dynamics simulations of simple liquids.  
Two temperature regimes can be introduced and interpreted as follow. At high 
temperature, the dynamics are well described by mode-coupling equations, in 
which the vertices are determined from the static correlations of the 
supercooled liquid.
This regime corresponds in our fits to a linear evolution of the mode-coupling
parameters with temperature.
Approaching the MCT critical transition, the effective 
vertices depart from the linear behavior and exhibit an unexpected behavior~: 
they follow the transition line without crossing it.
The avoidance of the transition has been known for a long time and is 
generally associated with the occurrence of so-called activated processes 
that are not included in the theory.
However, in our picture, these competitive processes seem to be
efficient well above the critical temperature predicted within the MCT
formalism, in a domain where activated or hopping processes are not supposed 
to be important.
An immediate consequence of this is that the universal, critical predictions 
of the theory should be unobservable experimentally.

That the MCT of the glass transition needs be completed is not a new idea,
but it is {\em a priori} a difficult task.
If the picture given here is valid, one would expect from a successful 
extension of MCT that it restores the quasi-linear variation of the
effective vertices by introducing new ones that
become important only at low temperature.
In this respect, note that our results for the critical parameters
extrapolated from high temperatures (where the necessary corrections to
ideal MCT are thought to be of minor importance) do not allow to recover 
those from the so-called "extended mode-coupling theory" that has been
proposed as an improvement of the ideal mode-coupling theory and was tested
in its scaling form on salol and CKN \cite{cum2}.

At last, it is necessary to point some remaining difficulties, that are of
great interest but are out of the scope of this study. 
Of critical importance is the nature of the relaxation probed by the 
depolarized light scattering, density fluctuations or molecular 
reorientations \cite{cum5}. 
Fortunately, we find that the choice for the 
expression of the scattered intensity as a function of the schematic 
correlators has no influence on the results of fit.
But it asks the question of the dynamics of molecular orientational processes
in the supercooled liquid phase and its role in the structural relaxation
of molecular systems. Some evidence exists in simulations that this dynamics 
is the same as for density fluctuations and that the two processes are 
coupled \cite{barrat2}. The very good results of the schematic fits are also in favor
of such a conclusion. Nevertheless, this remains to be demonstrated 
theoretically. Thus, in order to test the relevance of this approach from the 
experimental point of view, 
analysis of data from different spectroscopic techniques are necessary. 
With this goal, coherent neutron scattering spectra (that probe directly the 
dynamic structure factor) and far infrared absorption data (that probe
dipolar reorientations, thus molecular rotations) are now under study. 

\section*{Acknowledgments}
\indent

The authors are grateful to Prof.\ W. G\"{o}tze for very helpful comments
all along this work and Dr.\ M. Fuchs for providing them the codes for some 
parts of the calculations.
We are also indebted to Prof.\ H.Z. Cummins and Dr.\ G. Li
for the use of their experimental data on salol and CKN. 
We particularly acknowledge Dr.\ G. Tarjus for fruitful discussions and help in
our understanding of the theory.

\section*{Note}
\indent

After submission of this work, a paper was sent to us by W.
G\"{o}tze, which reports schematic calculations on DLS spectra of
glycerol\cite{schglyc}. The model used in this paper and the
results obtained will be discussed and
compared to ours in further publication \cite{soon}.

\clearpage
\begin{table}
\caption{Parameters for MCT fits to salol data ($T_{m}=315$ K, 
$T_{g}=218$ K).}
\vspace{.5 cm}
\begin{center}
\begin{minipage}{9.9cm}
\renewcommand{\footnoterule}{}
\begin{tabular}{||c|c|c|c||}   \hline
\label{salpast}
technique & $T_c$ (K) & $\lambda$ & references  \\ 
\hline \hline
DLS\footnote{depolarized light scattering}, ideal MCT & 256 $\pm$ 5 
& 0.70  & \citen{cum1}  \\ \hline
DLS$^{a}$, extended MCT & 250 & 0.73 &\citen{cum2} \\ \hline
neutron scattering & 263 $\pm$ 7 & & \citen{toul} \\ \hline
Brillouin scattering & 275 $\pm$ 10 & & \citen{mjl} \\ \hline
viscosity & 265 $\pm$ 5 & &\citen{cum1}  \\ \hline
ISLS\footnote{impulsive stimulated light scattering} & 266 $\pm$ 1 & &
\citen{yang} \\ \hline 
\end{tabular} 
\end{minipage}
\end{center}
\end{table}

\begin{table}
\caption{Parameters for MCT fits to CKN data ($T_{m}=433$ K, $T_{g}=333$ K).}
\vspace{.5 cm}
\begin{center}
\begin{minipage}{9.9cm}
\renewcommand{\footnoterule}{}
\begin{tabular}{||c|c|c|c||}   \hline
\label{cknpast}
technique & $T_c$ (K) & $\lambda$ & references  \\ \hline \hline
DLS\footnote{depolarized light scattering}, ideal MCT & 378 $\pm$ 5 & 0.81
& \citen{cum3}  \\ \hline
DLS$^{a}$, extended MCT & 378 & 0.85 & \citen{cum2} \\ \hline
neutron scattering & 368 $\pm$ 5 & & \citen{mezei} \\ \hline
Brillouin scattering & 375 $\pm$ 5 & &  \citen{cum4} \\ \hline
ISLS\footnote{impulsive stimulated light scattering} & 378 $\pm$ 2 & &
\citen{yang2} \\ \hline
Dielectric spectroscopy & 375 & 0.76 & \citen{lunken} \\ \hline
\end{tabular}
\end{minipage}
\end{center}
\end{table}
\clearpage
\begin{table}
\caption{Values of the temperature-independent parameters of the schematic 
model}
\vspace{.5 cm}
\begin{center}
\begin{minipage}{7.5cm}
\begin{center}
\begin{tabular}{||c|c|c||}   \hline
\label{global}
  & salol & CKN  \\ \hline \hline
$\Omega_{0}/2\,\pi$ (GHz) & 1230  & 2000 \\ \hline
$\Gamma_{0}$ & 0.11 & 0.3 \\ \hline
$\Omega_{1}/2\,\pi$ (GHz) & 160 & 900 \\ \hline
$\Gamma_{1}$ & 5.2 & 2.6 \\ \hline
$r$ & 37 & 10 \\ \hline
$\gamma$ & 18.6 & 2.4 \\ \hline
\end{tabular} 
\end{center}
\end{minipage}
\end{center}
\end{table}

\begin{table}
\caption{Parameters from the schematic-model fits to salol and CKN data}
\vspace{.5 cm}
\begin{center}
\begin{minipage}{7.5cm}
\begin{center}
\begin{tabular}{||c|c|c||}   \hline
\label{trans}
  & salol & CKN  \\ \hline \hline
$T_{c}$ (K) & 257 $\pm$ 5     & 388 $\pm$ 5 \\ \hline
$\lambda_{HT}$   & 0.69 $\pm$ 0.02 & 0.73 $\pm$ 0.02 \\ \hline
$\lambda_{vert}$   & 0.61 $\pm$ 0.02 & 0.69 $\pm$ 0.02 \\ \hline
\end{tabular} 
\end{center}
\end{minipage}
\end{center}
\end{table}

\clearpage
\begin{center}
FIGURE CAPTIONS \\~ \\~ \\
\end{center}

\begin{figlist}
\item \label{fit} Experimental depolarized light scattering spectra 
(for readability, some spectra are omitted and
the data are reported every 2 points) and schematic fits. 
a) salol (the last two curves at 233 K and 253 K are obtained with a 90 GHz
low frequency cut-off; the spectra at 243 K, 273 K, 293 K, 313 K and 333 K are
omitted).
b) CKN (the spectra at 383 K, 403 K, 423 K, 443 K and 468 K are omitted).   
\item \label{split} Individual contributions to the fitted DLS spectra of 
each of the two correlators of the schematic model in a semilogarithmic plot.
a) salol at 333 K.
b) CKN at 433 K.
Note that the $\chi_{0}"$ contribution is negligible below
50 GHz for salol, but not for CKN.
This explains the equivalence between a scaling-equation fit and our 
schematic calculations in the case of salol.
\item \label{v1v2T} Vertices $v_{1}$ and $v_{2}$ of the schematic model as
functions of temperature for salol and CKN. The straight lines correspond
to the linear extrapolation described in the text.
\item \label{v1v2} Two-dimensional ($v_1, v_2$) parameter space of the 
schematic mo\-del and trajectories followed by salol and CKN with temperature.
\item \label{lambda} Effective and local exponent parameters (resp. 
$\lambda_{e\!f\!f}$ and $\lambda_{loc}$) as functions of $\varepsilon$ 
(see text for definition) for the one-correlator $F_{12}$-model. 
\item \label{lambsalol} Effective temperature-dependent, effective  
temperature-indep\-en\-dent, and local exponent parameters (resp. 
$\lambda_{e\!f\!f}$, $\lambda_{global}$, and $\lambda_{loc}$) as functions of
temperature for supercooled salol above $T_{c}$.
\item \label{split2} Individual contributions to the fitted DLS spectra of 
each of the two correlators of the schematic model at the closest temperatures
higher than $T_{c}$ (the scaling factor is such that the minima of 
susceptibilities are equal).
a) salol at 263 K.
b) CKN at 393 K.
Note that the minima of both contributions (when they exist) occur at 
different frequencies, in disagreement with the asymptotic predictions of
the mode-coupling theory.
\end{figlist}
%\setlength{\textheight}{22cm}

%\clearpage
%\thispagestyle{empty}
%\begin{figure}
%\scalebox{1.2}{\includegraphics[5cm,5cm][10cm,22.5cm]{fig1.ps}}
%\end{figure}

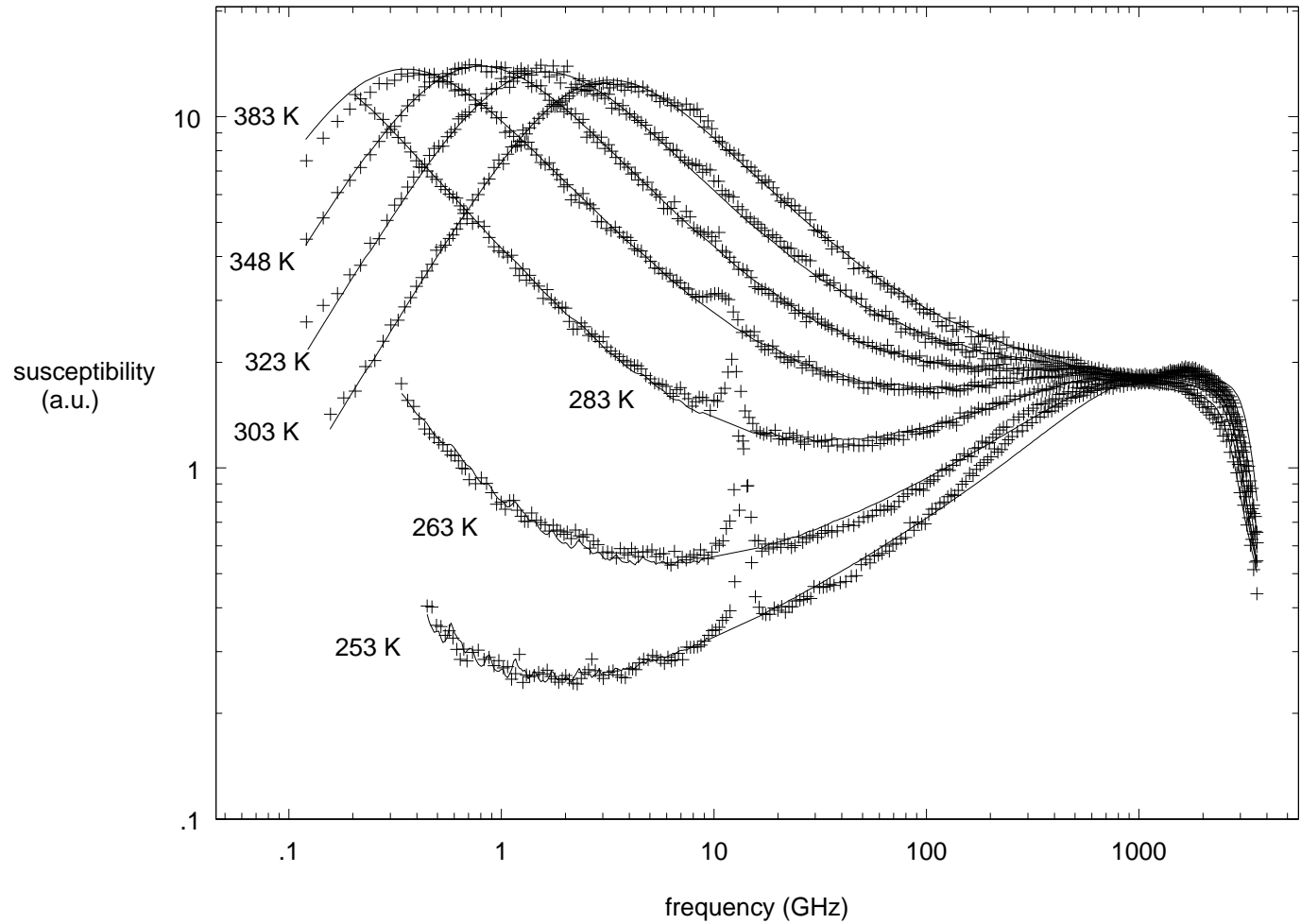
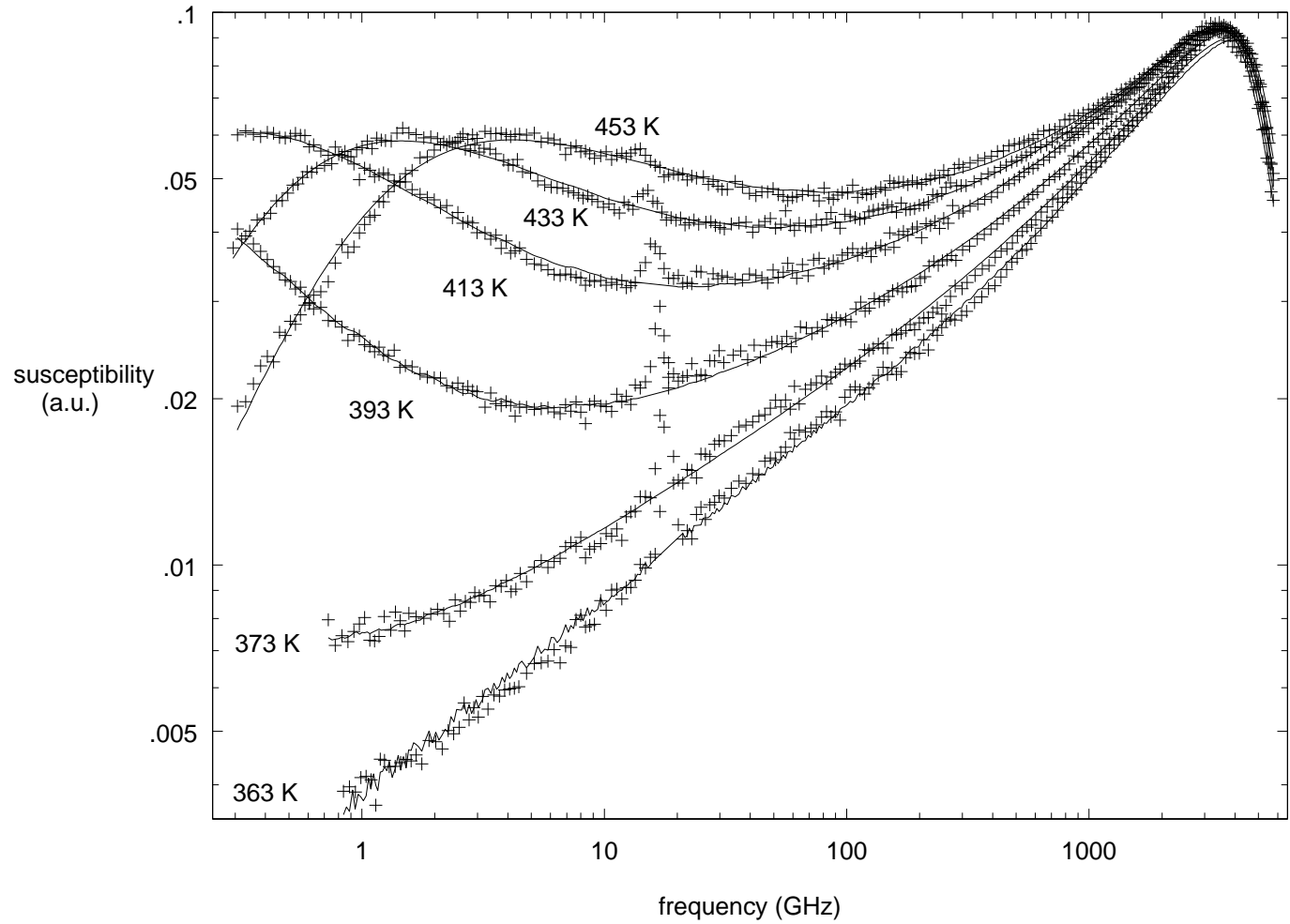
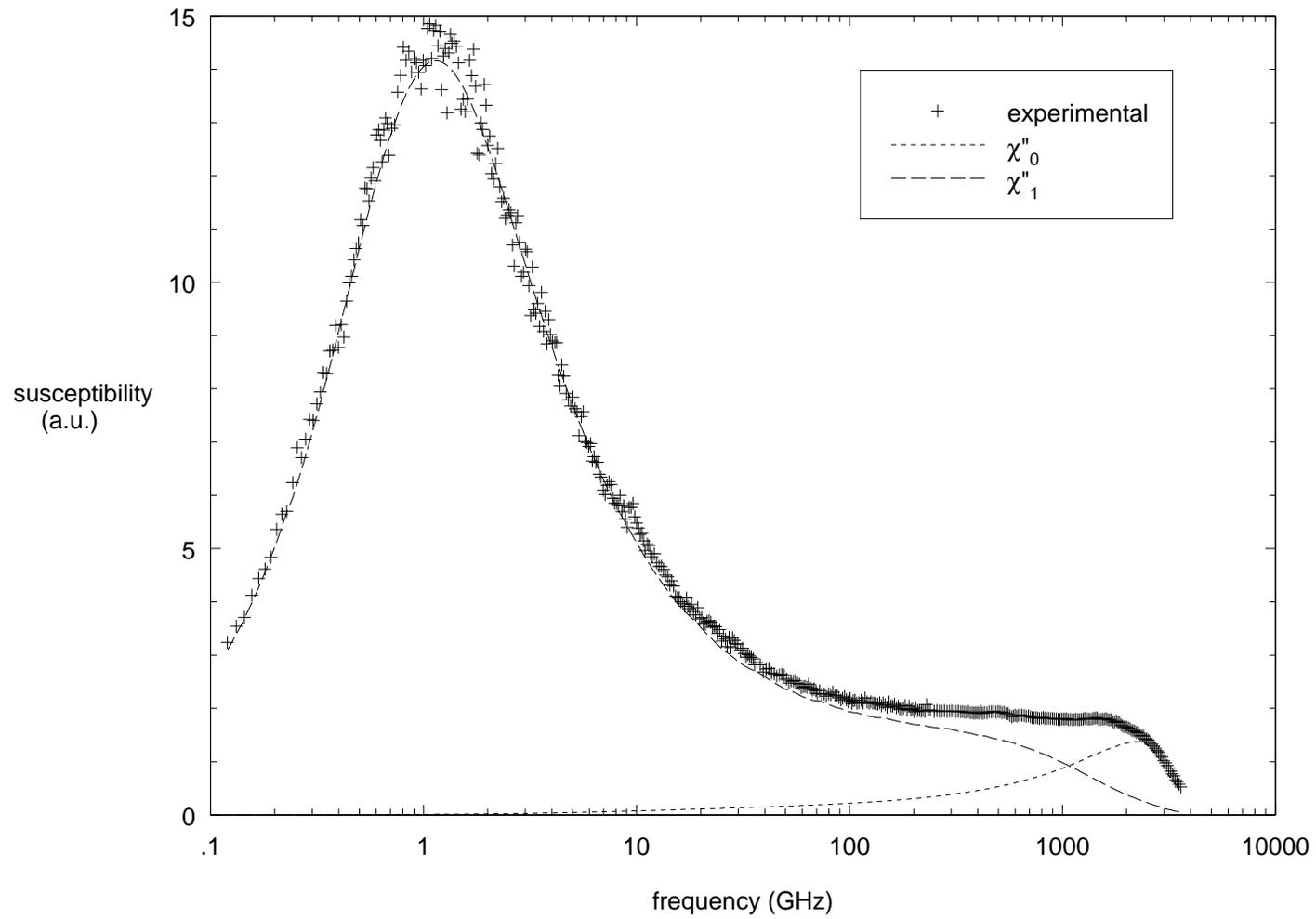
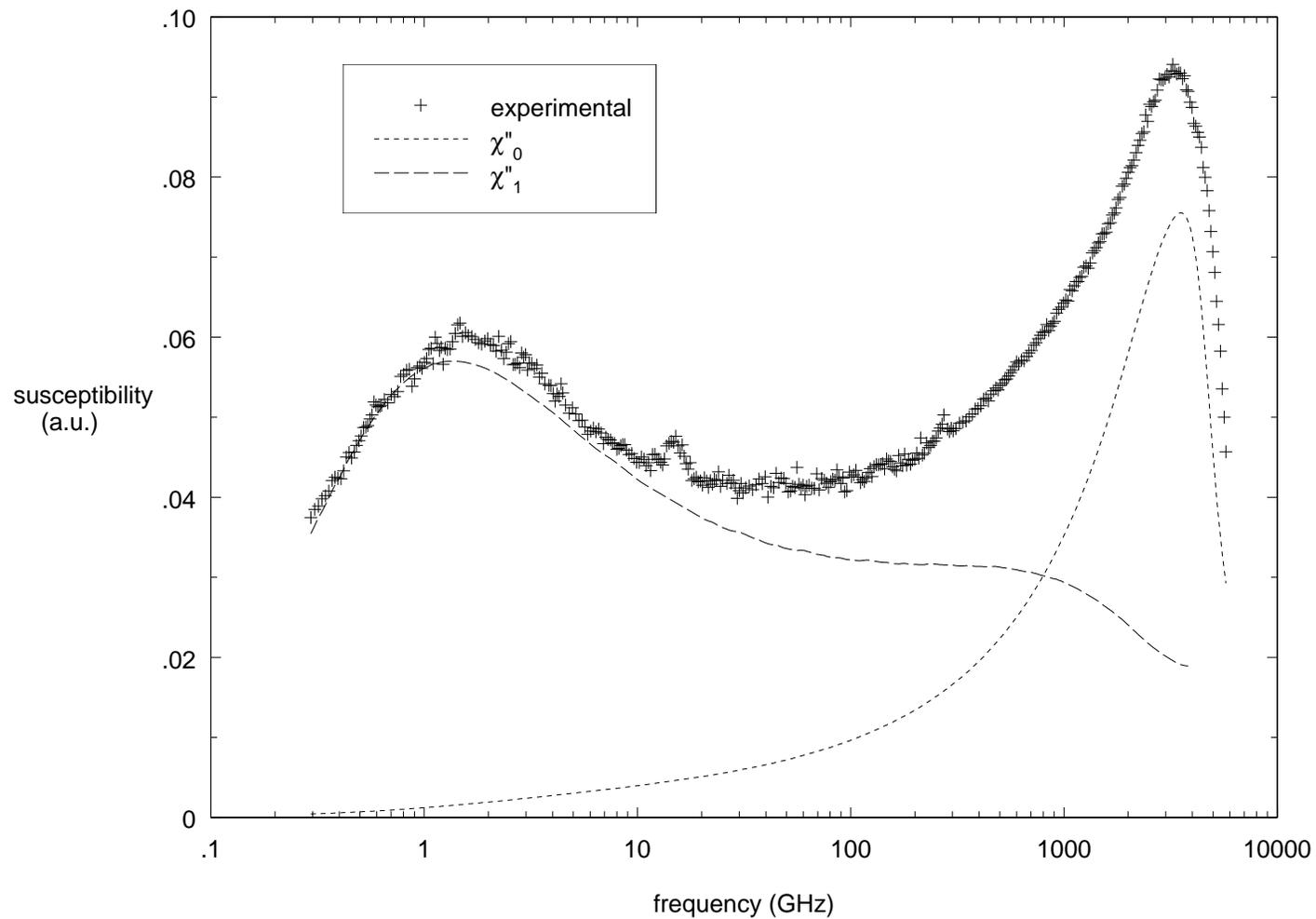
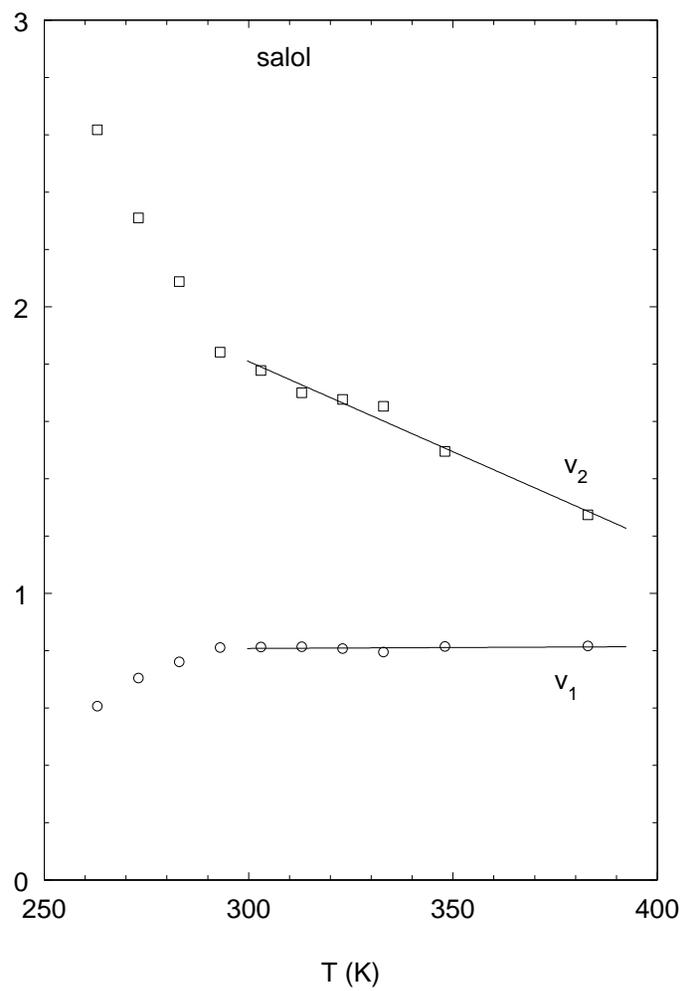
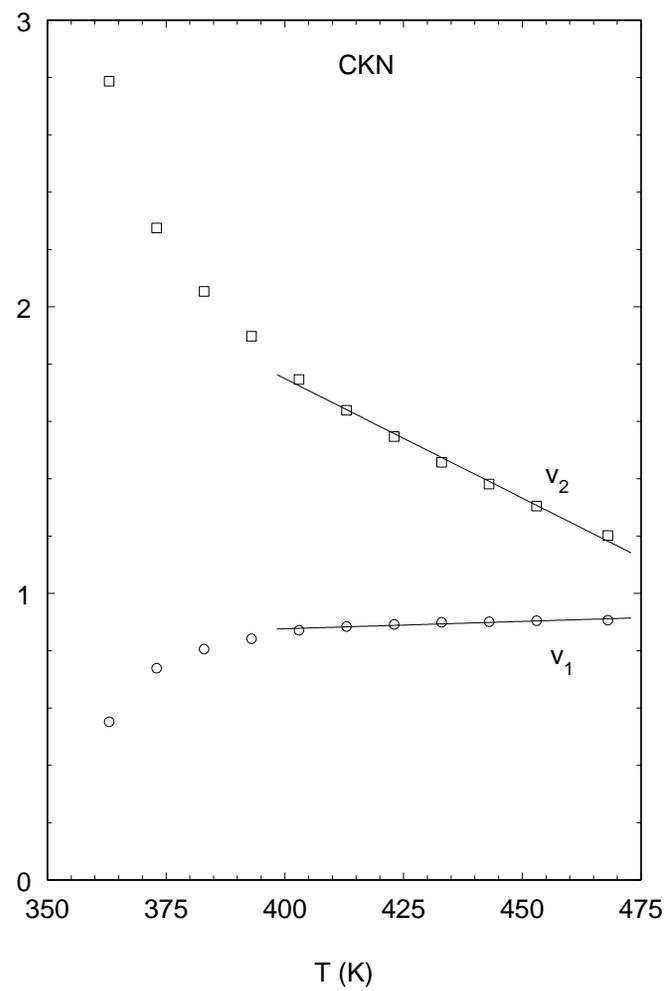
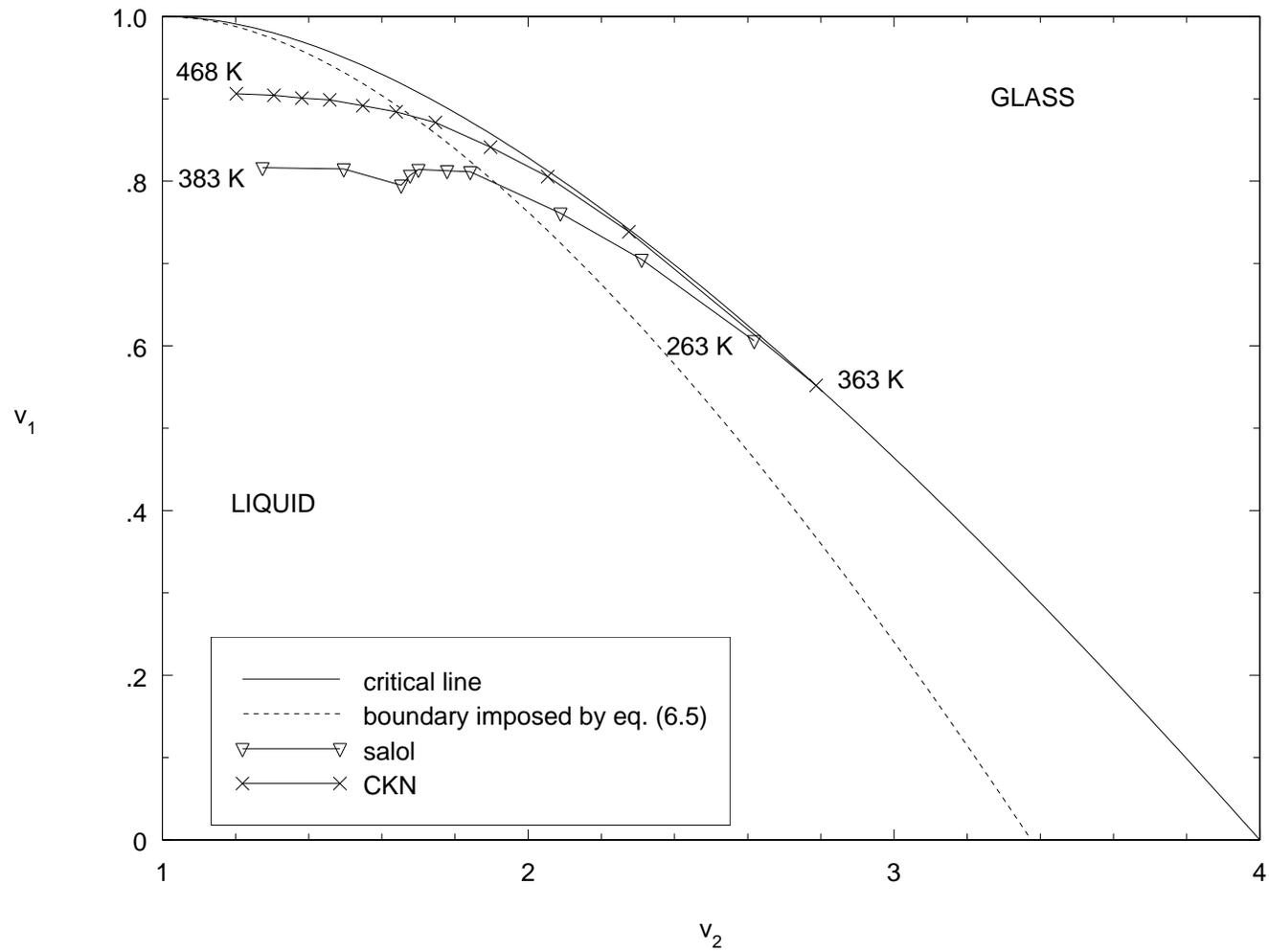
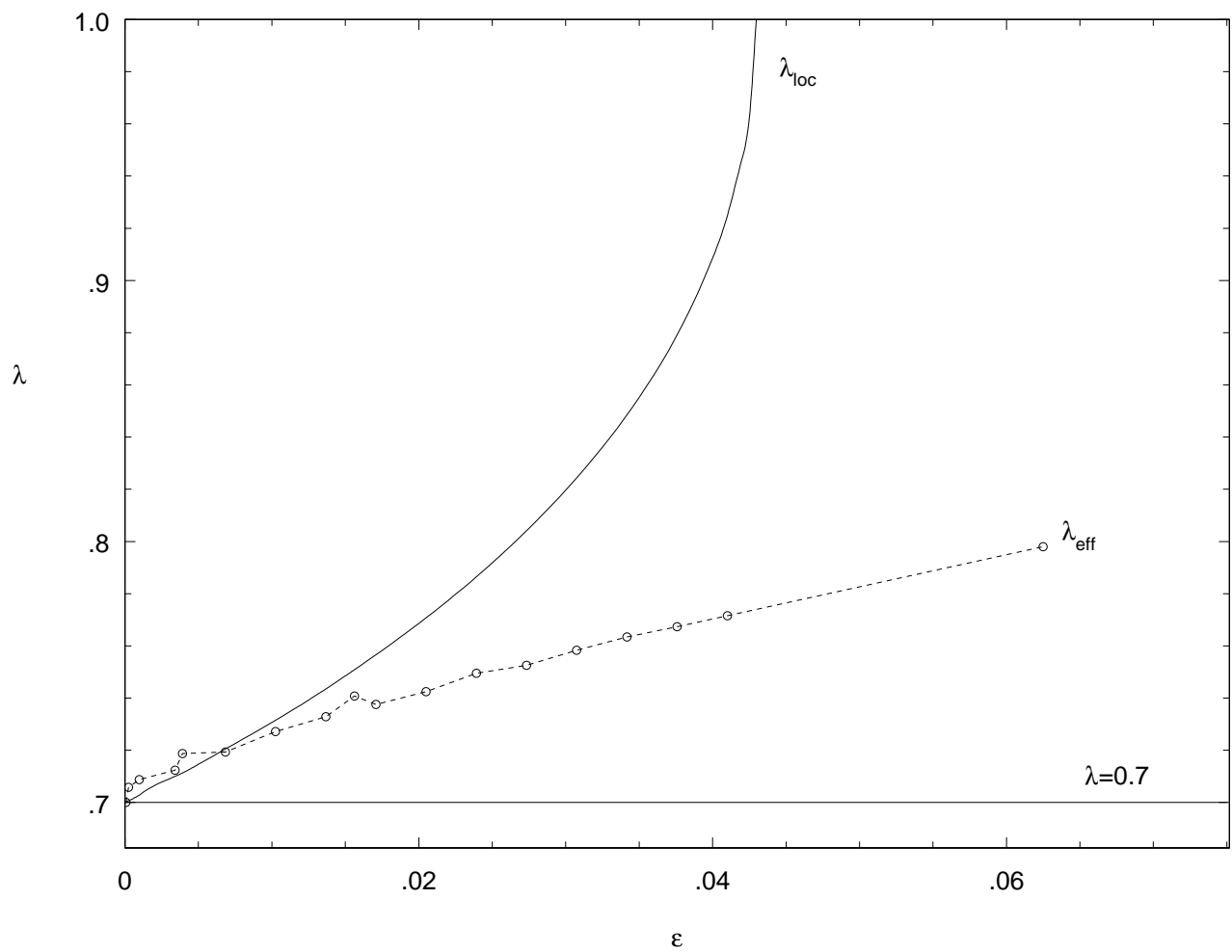
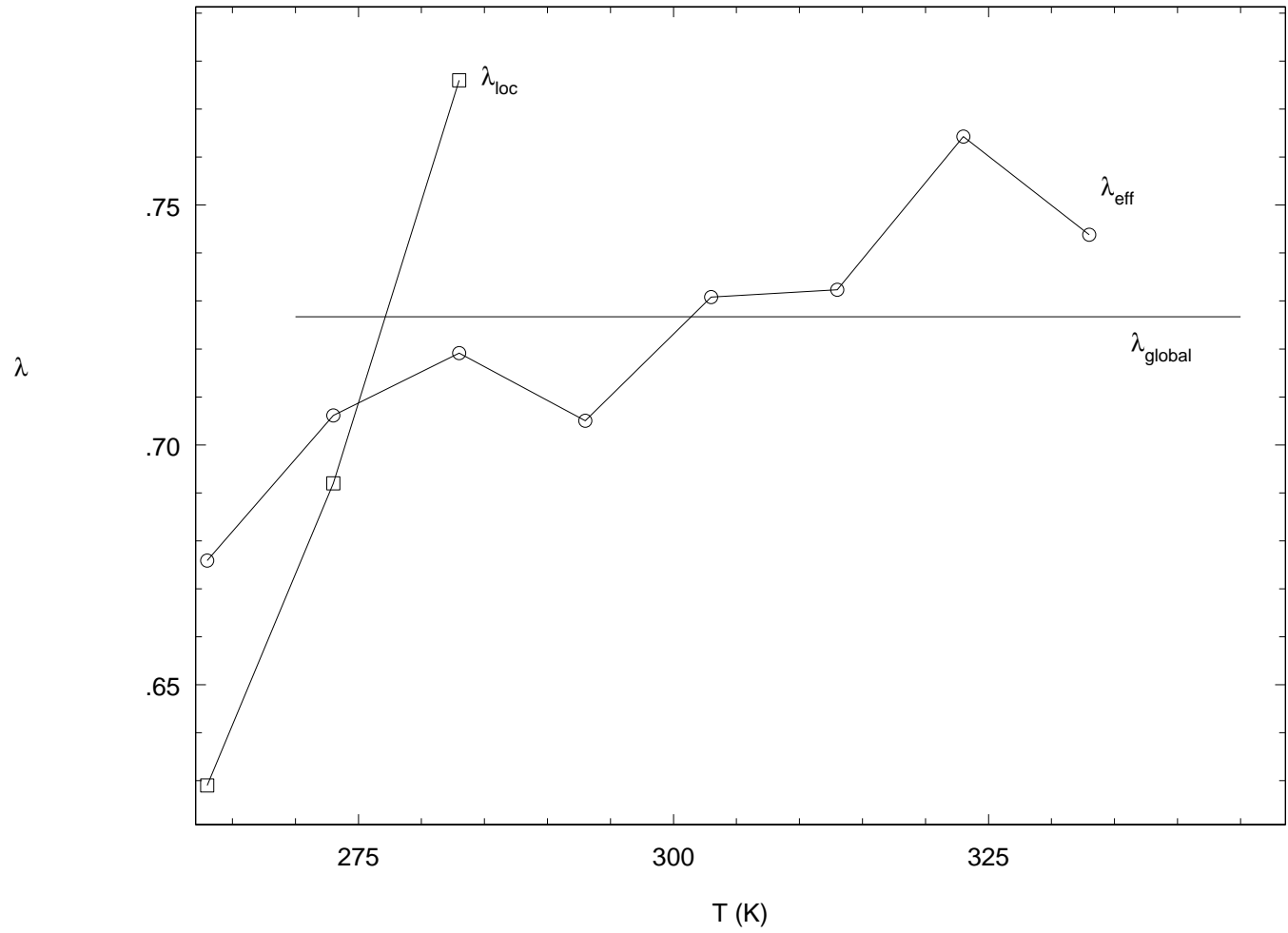
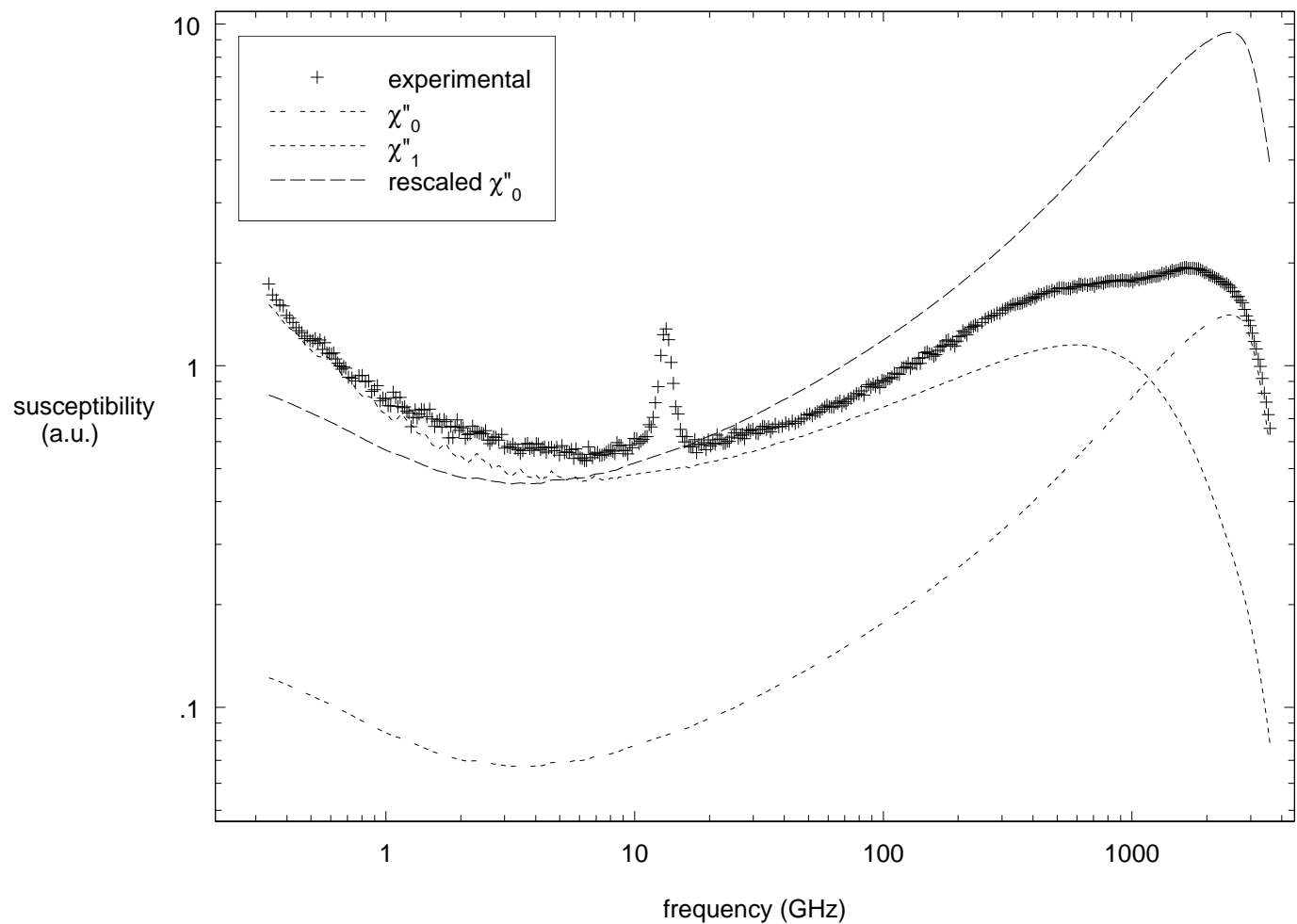
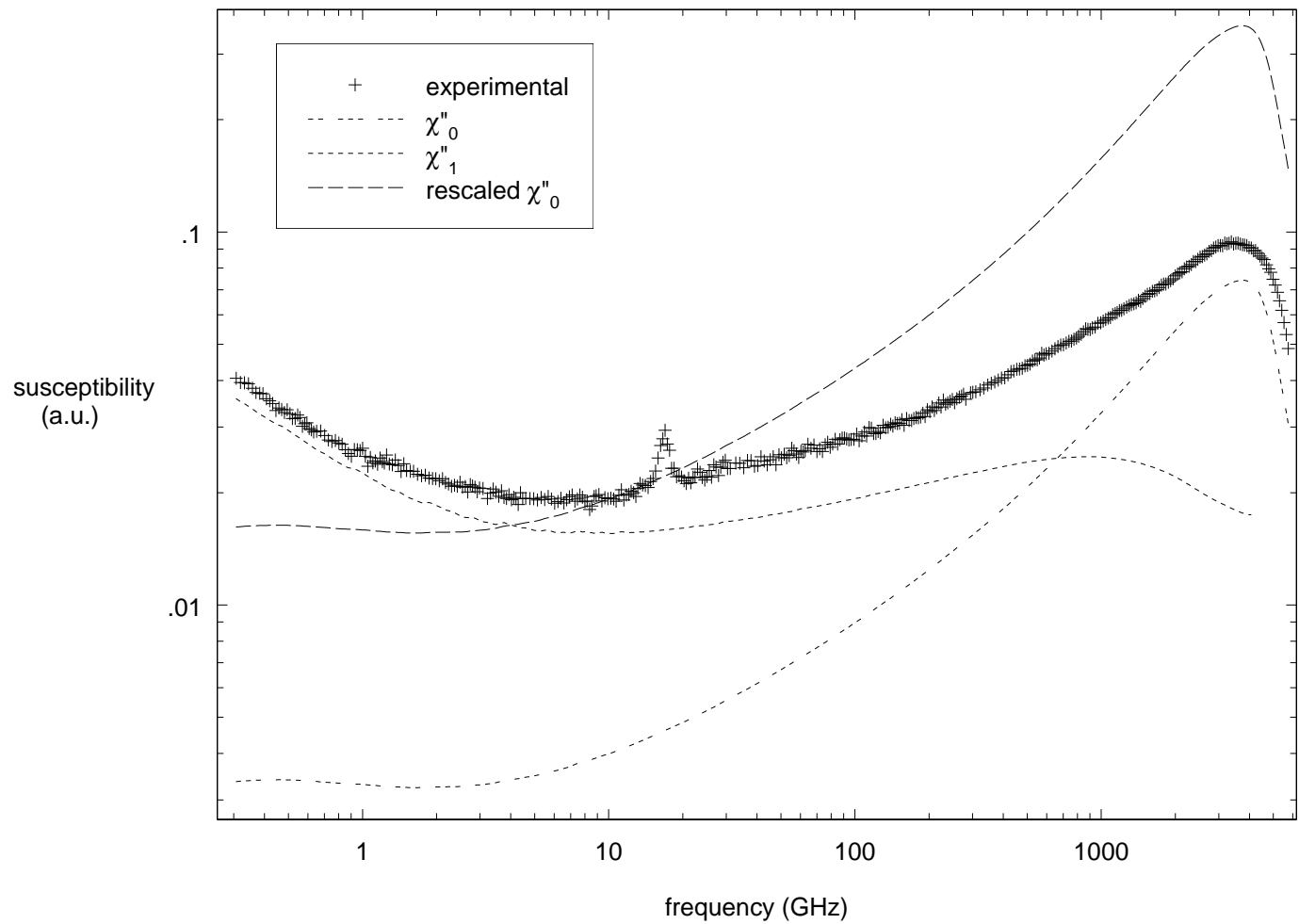
\end{document}